\documentclass[11pt, a4paper]{article}
\pdfoutput=1 

\usepackage[height=8.85in,width=6.75in]{geometry}

\usepackage{graphicx,rotating}     
\usepackage[bookmarksopen,colorlinks=true,linkcolor=steelblue,
citecolor=orangered,urlcolor=darkred,linktoc=all]{hyperref}

\usepackage{amsmath,amssymb,color,xcolor}
\usepackage{slashed}
\usepackage{bm}
\usepackage{bbm}
\usepackage{cite}
\usepackage{comment}
\usepackage[utf8]{inputenc}
\usepackage{accents}
\usepackage[footnotesize]{caption}
\usepackage{cancel}

\newcommand{\abs}[1]{\left\vert {#1} \right\vert}
\newcommand{\romantwo}{I\hspace{-0.2mm}I}

\makeatletter
\@addtoreset{equation}{section}

\makeatother

\usepackage{xcolor}
\definecolor{darkred}{rgb}{0.7, 0., 0.}
\definecolor{orangered}{rgb}{1,0.27,0.}
\definecolor{steelblue}{rgb}{0.275,0.51, 0.706}
\definecolor{forestgreen}{rgb}{0.13,0.55,0.13}

\usepackage{tikz}
\usepackage{tikz-feynman}
\tikzfeynmanset{compat=1.0.0}
\pgfdeclarelayer{bg}    
\pgfsetlayers{bg,main}  

\begin{document}

\hypersetup{pageanchor=false}
\begin{titlepage}

\begin{center}

\hfill UMN-TH-4217/23 \\
\hfill FTPI-MINN-23-10 \\

\vskip 0.5in

{\Huge \bfseries Heavy Neutral Leptons \vspace{5mm} \\ from Stopped Muons and Pions
} \\
\vskip .8in

{\Large Yohei Ema,$^{a,b}$ Zhen Liu,$^{b}$ Kun-Feng Lyu,$^{b}$ Maxim Pospelov$^{a,b}$}

\vskip .3in
\begin{tabular}{ll}
$^a$
&\!\!\!\!\!\emph{William I. Fine Theoretical Physics Institute, School of Physics and Astronomy,}\\
&\!\!\!\!\!\emph{University of Minnesota, Minneapolis, MN 55455, USA} \\
$^b$
&\!\!\!\!\!\emph{School of Physics and Astronomy, University of Minnesota, Minneapolis, MN 55455, USA}
\end{tabular}

\end{center}
\vskip .6in

\begin{abstract}
\noindent
Stopped muons, which are generic in pion-at-rest experiments, can shed light on heavy neutral leptons (HNLs) 
in unexplored parameter spaces. 
If the HNL is lighter than the muon, the HNL can be produced from decays of muons and pions.
The HNL will travel from the production location and decay into visible Standard Model (SM) modes, 
leaving signals inside downstream detectors.
We find that in the case that the HNL dominantly mixes with muon neutrinos, 
the LSND constraint on the mixing angle squared is stronger than all the previous constraints
by more than an order of magnitude. In this study, we recast the LSND measurement of the $\nu-e$ scattering. 
Future experiments such as PIP2-BD could further improve the sensitivity, provided they can distinguish the HNL events from backgrounds induced by the SM neutrinos.
\end{abstract}

\end{titlepage}

\tableofcontents
\renewcommand{\thepage}{\arabic{page}}
\renewcommand{\thefootnote}{$\natural$\arabic{footnote}}
\setcounter{footnote}{0}
\hypersetup{pageanchor=true}

\section{Introduction}
\label{sec:introduction}

The discovery of neutrino oscillation stands out as clear evidence of physics beyond the 
Standard Model (SM)~\cite{Davis:1968cp,Super-Kamiokande:1998kpq,SNO:2002tuh,Esteban:2020cvm}.
Neutrinos are exactly massless within the SM, and hence non-zero neutrino masses require an extension of the SM in the neutrino sector.

Among possible extensions, the introduction of heavy neutral leptons (HNLs) is particularly interesting due to its minimality
(see e.g., Refs~\cite{Boyarsky:2009ix,Drewes:2013gca,Dasgupta:2021ies,Abdullahi:2022jlv,Batell:2022xau} for representative reviews of the HNL).
The HNL can couple to the SM lepton and Higgs doublets at the renormalizable level. 
After the electroweak symmetry breaking, the HNL interacts with the SM particles through mixing with the SM neutrinos.
The phenomenology of the HNL can be diverse, depending on the mass of the HNL.
If the HNL is light, $m_N \lesssim 1\,\mathrm{eV}$ where $m_N$ is the HNL mass,
the HNL can participate in the neutrino oscillation, which may be relevant for, \emph{e.g.}, 
the anomalies observed at LSND~\cite{LSND:2001aii} and MiniBooNE~\cite{MiniBooNE:2018esg}.

Even if the HNL is heavy enough to be irrelevant for the neutrino oscillation, 
it can still leave signals in different types of experiments.
In particular, if the HNL is lighter than the muon,
the HNL can be produced from the decay of muons and pions.
Once produced, the HNL can travel downwards and decay into an $e^+e^-$ pair and a neutrino,
leaving signals inside downstream detectors, as long as $m_N > 2m_e$ with $m_e$ being the electron mass.

Typically, one can search for HNLs in the particle regime in the high-energy beamdump experiments and as well at colliders. In the regime of HNL mass around 10-100~MeV, the searches are dominated by HNL from pion decays and kaon decays. In this paper, we point out that stopped muons, which commonly exist in the pion-at-rest experiments, can provide leading productions and enables us to probe new parameter spaces in HNLs. 
In this paper, we focus on neutrino experiments with stopped muons and stopped pions as the source of the HNL, such as
LSND and PIP2-BD~\cite{Pellico:2022dju,Toups:2022yxs}.
In the past, the LSND experiment has been used to set constraints on dark photons, dark scalars, and light dark matter (see {\em, e.g.,} \cite{Batell:2009di,deNiverville:2011it,Foroughi-Abari:2020gju,Ema:2022afm}).
In the parameter region of our interest, the HNL decay length is far longer than the typical size of the laboratory.
The small velocity of the HNL from the muon and pion decay-at-rest, in comparison to that decay-in-flight, therefore
enhances the probability of the HNL decaying inside the detectors, strengthening the sensitivity.
Together with the large accumulation of proton-on-target (POT), 
we find that the LSND experiment provides an upper limit on the mixing angle
between the muon neutrino and the HNL which is stronger by more than an order of magnitude than
the previous constraints from T2K~\cite{T2K:2019jwa,Arguelles:2021dqn} and 
$\mu$BooNE~\cite{MicroBooNE:2021usw,Kelly:2021xbv}.
Interestingly, our LSND constraint still does not utilize all the properties of the HNL decay, as LSND did not perform a search of the $e^+ e^-$ events and thus the HNL events are compared with the single $e$ events originating from the SM neutrinos.
Future experiments such as PIP2-BD potentially improve the sensitivity significantly, provided that they can distinguish
the HNL-induced $e^+ e^-$ events from backgrounds induced by, \emph{e.g.}, the SM neutrinos. While this part of the parameter space is disfavored by big bang nucleosynthesis (BBN) in the {\em minimal} model of HNLs, a trivial model modification, such as dark decay channels can be introduced to render this model safe from the cosmological constraints.

The rest of this paper is organized as follows. In Sec.~\ref{sec:preliminary}, we summarize the production and decay rates
of the HNL. This section provides basic formulas that we use to derive the sensitivity in the subsequent section. 
We consider both the muon mixing and electron mixing cases, with an emphasis on the former case. 
In Sec.~\ref{sec:constraint}, we derive the current constraint on the mixing angle from LSND
and the future sensitivity of PIP2-BD. We summarize our results in Sec.~\ref{sec:summary}.

\section{Preliminary}
\label{sec:preliminary}

In this section, we summarize the basic ingredients necessary for our analysis.
For definiteness, we consider one (pseudo-)Dirac HNL which dominantly mixes 
with either muon or electron neutrinos.
The relevant part of the Lagrangian is given by
\begin{align}
	\mathcal{L}
	&= \bar{N}\left(i\slashed{\partial} -m_N\right)N
	- \frac{g}{\sqrt{2}}U_{lN}^*\bar{l} \slashed{W}^- N
	- \frac{g}{2\cos\theta_W} U_{lN}^*\bar{\nu}_l \slashed{Z} N
	+ (\mathrm{h.c.}),
\end{align}
where $N$ is the HNL field with its mass $m_N$, $g$ is the SU(2) gauge coupling, $\theta_W$ is the weak mixing angle, 
and $U_{lN}$ is the mixing angle between the HNL and the SM neutrino.
We mainly focus on the case $m_N < m_\mu$, where $m_\mu$ is the muon mass, so that the HNL can be produced by the decay of (stopped) muons and pions.
We consider both the muon mixing case $l = \mu$ and the electron mixing case $l = e$,
with a particular interest in the former case.
Throughout this paper, we ignore the electron and neutrino masses as the energy scale of our interest 
is above $\sim 10\,\mathrm{MeV}$.

Throughout this paper, we assume (pseudo-)Dirac mass $m_N$. While in principle, both Dirac and Majorana masses are possible, the (pseudo-)Dirac mass is easier to reconcile with neutrino phenomenology. In the inverse seesaw models, the HNLs are dominated by their Dirac mass and receive a small correction from their Majorana mass terms~\footnote{hence dubbed pseudo-Dirac.}.  A single generation with Majorana mass would imply $|U_{l N}|^2m_N$ contributions to the active neutrino masses, which is above the neutrino mass limits for the interesting values of these parameters, implying that the Majorana mass $m_N$ would require an additional finely tuned contribution to $m_\nu$. In general multi-generation HNLs with all mixings with different neutrinos turned on, an inverse seesaw-like mass spectrum is generically found to be consistent with the SM neutrino mass considerations, see, e.g., discussions in review~\cite{Abdullahi:2022jlv}.

\subsection{HNL production rate}
\label{subsec:production}

We first discuss the HNL production rates from the decay of muons and pions.
The muon mixing case and electron mixing case are studied separately.

\subsubsection*{Muon mixing case}
If $m_N < m_\mu$, 
the HNL is predominantly produced from the decay of muons
at neutrino experiments with stopped muons such as LSND.
The amplitude is diagrammatically given by
\begin{align}
	i\mathcal{M}(\mu \to e \nu_e N) = 
	\begin{tikzpicture}[baseline=(b)]
	\begin{feynman}[inline = (base.b), horizontal=b to d]
		\vertex [label=180:\({\scriptstyle \mu}\)](a);
		\vertex [right = of a] (b);
		\vertex [above right = of b, label=360:\({\scriptstyle N}\)] (c);
		\vertex [below right = of b] (d);
		\vertex [above right = of d, label=360:\({\scriptstyle e}\)] (e);
		\vertex [below right = of d, label=360:\({\scriptstyle \nu_e}\)] (f);
		\diagram*{
		(a) -- [fermion] (b) -- [fermion] (c),
		(b) -- [photon] (d),
		(f) -- [fermion] (d) -- [fermion] (e),
		};
	\end{feynman}
	\end{tikzpicture}.
\end{align}
The decay rate is expressed as
\begin{align}
	\Gamma(\mu \to e\nu_e N) =
	\int_{m_N}^{(m_\mu^2+m_N^2)/2m_\mu}dE_N \frac{d\Gamma(\mu\to Ne\nu_e)}{dE_N},
\end{align}
with the differential decay rate given by
\begin{align}
	\frac{d\Gamma(\mu \to e\nu_e N)}{dE_N} = 
	\frac{G_F^2 \abs{U_{\mu N}}^2}{12\pi^3}
	\,\left(3E_N(m_\mu^2 + m_N^2) - 4m_\mu E_N^2 - 2m_\mu m_N^2 \right)
	\sqrt{E_N^2 - m_N^2},
\end{align}
where $G_F$ is the Fermi constant.
If we set $m_N = 0$, this formula correctly reproduces the well-known muon decay rate 
up to the factor $\vert U_{\mu N}\vert^2$.

Although {\it subdominant}, if $m_N < m_\pi - m_\mu$ with $m_\pi$ the pion mass,
the HNL can also be produced from the decay of pions,
whose diagram is given by
\begin{align}
	i\mathcal{M}(\pi \to \mu N)
	&= \begin{tikzpicture}[baseline=(b)]
	\begin{feynman}[inline = (base.b), horizontal=b to d]
		\vertex [label=180:\({\scriptstyle \pi}\)](a);
		\vertex [right = of a] (b);
		\vertex [above right = of b, label=360:\({\scriptstyle N}\)] (c);
		\vertex [below right = of b, label=360:\({\scriptstyle \mu}\)] (d);
		\diagram*{
		(a) -- [scalar] (b),
		(c) -- [fermion] (b) -- [fermion] (d),
		};
	\end{feynman}
	\end{tikzpicture}.
\end{align}
In this case, the final-state HNL is monochromatic, and the differential decay rate is given by
\begin{align}
	\frac{d\Gamma(\pi \to \mu N)}{dE_N} &= 
	\frac{G_F^2 f_\pi^2 \abs{V_{ud}}^2 \abs{U_{\mu N}}^2}{8\pi m_\pi^3}
	\left((m_\mu^2 + m_N^2) m_\pi^2 - (m_\mu^2 + m_N^2)^2 + 4m_\mu^2 m_N^2\right)
	\nonumber \\
	&\times
	\sqrt{m_\pi^4 - 2(m_\mu^2 + m_N^2)m_\pi^2 + (m_\mu^2 - m_N^2)^2}
	\times \delta\left(E_N - \frac{m_\pi^2 + m_N^2 - m_\mu^2}{2m_\pi}\right),
\end{align}
where $V_{ud}$ is the $ud$-component of the CKM matrix and $f_\pi$ is the pion decay constant.

\subsubsection*{Electron mixing case}

As in the muon mixing case, the HNL can be produced from the muon decay if $m_N < m_\mu$, 
and the relevant amplitude is given diagrammatically as
\begin{align}
	i\mathcal{M}(\mu \to e \nu_\mu N) = 
	\begin{tikzpicture}[baseline=(b)]
	\begin{feynman}[inline = (base.b), horizontal=b to d]
		\vertex [label=180:\({\scriptstyle \mu}\)](a);
		\vertex [right = of a] (b);
		\vertex [above right = of b, label=360:\({\scriptstyle \nu_\mu}\)] (c);
		\vertex [below right = of b] (d);
		\vertex [above right = of d, label=360:\({\scriptstyle e}\)] (e);
		\vertex [below right = of d, label=360:\({\scriptstyle N}\)] (f);
		\diagram*{
		(a) -- [fermion] (b) -- [fermion] (c),
		(b) -- [photon] (d),
		(f) -- [fermion] (d) -- [fermion] (e),
		};
	\end{feynman}
	\end{tikzpicture}.
\end{align}
The decay rate is given by
\begin{align}
	\Gamma(\mu \to e \nu_\mu N) = \int_{m_N}^{(m_\mu^2 + m_N^2)/2m_\mu}
	dE_N\frac{d\Gamma(\mu \to e \nu_\mu N)}{dE_N},
\end{align}
with the differential rate given by
\begin{align}
	\frac{d\Gamma(\mu \to e \nu_\mu N)}{dE_N} &=
	\frac{G_F^2 \abs{U_{eN}}^2}{2\pi^3}E_N(m_\mu^2 + m_N^2 - 2m_\mu E_N)\sqrt{E_N^2 - m_N^2}.
\end{align}
This formula again correctly reproduces the well-known muon decay rate in the limit $m_N = 0$.

In the electron mixing case, the pion produces the HNL if $m_N < m_\pi$.
The contribution of the pion decay is important at the relatively high $m_N$ region.
This is in contrast to the muon mixing case, where the decay is kinematically allowed only for $m_N < m_\pi - m_\mu$.
The decay rate of pion is obtained simply by replacing $\mu \to e$ in the muon mixing case as
\begin{align}
	\frac{d\Gamma(\pi \to eN)}{dE_N} &= 
	\frac{G_F^2 f_\pi^2 \abs{V_{ud}}^2 \abs{U_{eN}}^2 m_N^2}{8\pi m_\pi^3}
	\left(m_\pi^2 - m_N^2\right)^2
	\times \delta\left(E_N - \frac{m_\pi^2 + m_N^2}{2m_\pi}\right).
\end{align}
Note that the chirality flip is supplied by the HNL and this rate is not suppressed by the electron mass.

\subsection{HNL decay rate}
\label{subsec:decay}
We next discuss the decay rate of the HNL.
We first note that, in the case of our interest, the total decay rate of the HNL $\Gamma_N$
into the SM particles is estimated as
\begin{align}
	\Gamma_N \sim \Gamma_\mu \left(\frac{m_N}{m_\mu}\right)^5 \abs{U_{l N}}^2,
	\label{eq:HNLdecay_estimation}
\end{align}
and thus the lifetime is estimated as
\begin{align}
	c \tau_N \sim 10^8\,\mathrm{m}\times  
	\left(\frac{m_\mu}{m_N}\right)^5 \left(\frac{10^{-6}}{\vert{U_{l N}}\vert^2}\right).
\end{align}
Thus $N$ is sufficiently long-lived at the laboratory scale and only a small fraction of the HNL decays
within the laboratory,
even after including the velocity of the HNL (smaller than the speed of light $c$).
Therefore we focus on the partial decay rate $\Gamma(N \to e^+ e^- \nu)$ 
as the other decay modes with only neutrinos in the final state are simply invisible.

\subsubsection*{Muon mixing case}

In the muon mixing case, the HNL decays via the neutral current.
The relevant diagram is given by
\begin{align}
	i\mathcal{M}(N \to e^+ e^- \nu_\mu) &= 
	\begin{tikzpicture}[baseline=(b)]
	\begin{feynman}[inline = (base.b), horizontal=b to d]
		\vertex [label=180:\({\scriptstyle N,\,p_N}\)](a);
		\vertex [right = of a] (b);
		\vertex [above right = of b, label=360:\({\scriptstyle \nu_\mu,\,p_\nu}\)] (c);
		\vertex [below right = of b] (d);
		\vertex [above right = of d, label=360:\({\scriptstyle e^-,\,p_2}\)] (e);
		\vertex [below right = of d, label=360:\({\scriptstyle e^+,\,p_1}\)] (f);
		\diagram*{
		(a) -- [fermion] (b) -- [fermion] (c),
		(b) -- [photon] (d),
		(f) -- [fermion] (d) -- [fermion] (e),
		};
	\end{feynman}
	\end{tikzpicture}.
\end{align}
The spin-averaged matrix element square is computed as
\begin{align}
	\abs{\mathcal{M}(N \to e^+ e^- \nu_\mu)}^2 = 16 G_F^2 \abs{U_{\mu N}}^2
	\left[4 \sin^4\theta_W (p_2 \cdot p_N) (p_1 \cdot p_\nu)
	+ \left(1-2\sin^2\theta_W\right)^2 (p_2 \cdot p_\nu) (p_1 \cdot p_N)
	\right],
\end{align}
and the decay rate is given by
\begin{align}
	\Gamma(N \to e^+ e^- \nu_\mu)
	&= 
	\frac{1}{(2\pi)^5}\frac{1}{64 m_N^3}\int_0^{m_N^2} dm_{2\nu}^2
	\int_{0}^{m_N^2 - m_{2\nu}^2} dm_{1\nu}^2
	\int_0^{2\pi}d\psi \int_{-1}^{1}d\cos\theta \int_{0}^{2\pi}d\phi \abs{\mathcal{M}}^2.
	\label{eq:decay_distribution_muon}
\end{align}
Here we have defined
\begin{align}
	m_{1\nu}^2 = (p_1 + p_\nu)^2 = m_N^2 - 2 m_N E_-^{N},
	\quad
	m_{2\nu}^2 = (p_2 + p_\nu)^2 = m_N^2 - 2 m_N E_+^{N},
\end{align}
where $E_\pm^{N}$ is the energy of $e^\pm$ in the $N$-rest frame,
and $(\psi, \theta, \phi)$ are the Euler angle of the orientation of the final particles 
(they are on a single plane) relative to the initial $N$.
Note that the angle between $\vec{p}_1$ and $\vec{p}_2$ in the $N$-rest frame is fixed once 
we fix $E_\pm^{N}$ due to the energy-momentum conservation as
\begin{align}
	\frac{\vec{p}_1 \cdot\vec{p}_2}{\vert \vec{p}_1\vert \vert \vec{p}_2\vert} 
	= 1 - \frac{m_N}{E_+^{N} E_-^{N}}\left(E_+^{N} + E_-^{N} - \frac{m_N}{2}\right).
\end{align}
Later we use this expression for the Monte Carlo (MC) simulation of the cut efficiency at the LSND experiment.
The total decay rate is easily obtained from this expression as
\begin{align}
	\Gamma(N \to e^+ e^- \nu_\mu) = \frac{G_F^2 m_N^5}{768\pi^3}\abs{U_{\mu N}}^2
	\left(1-4\sin^2\theta_W + 8\sin^4\theta_W\right).
\end{align}
This correctly reproduces the parameter dependence of Eq.~\eqref{eq:HNLdecay_estimation}.

\subsubsection*{Electron mixing case}

In the electron mixing case, the HNL can decay into $e^+e^- \nu_e$ both through the neutral and charged currents.
The amplitude is given by
\begin{align}
	i\mathcal{M}(N \to e^+ e^- \nu_e) &= 
	\begin{tikzpicture}[baseline=(b)]
	\begin{feynman}[inline = (base.b), horizontal=b to d]
		\vertex [label=180:\({\scriptstyle N,\,p_N}\)](a);
		\vertex [right = of a] (b);
		\vertex [above right = of b, label=360:\({\scriptstyle \nu_e,\,p_\nu}\)] (c);
		\vertex [below right = of b] (d);
		\vertex [above right = of d, label=360:\({\scriptstyle e^-,\,p_2}\)] (e);
		\vertex [below right = of d, label=360:\({\scriptstyle e^+,\,p_1}\)] (f);
		\diagram*{
		(a) -- [fermion] (b) -- [fermion] (c),
		(b) -- [photon] (d),
		(f) -- [fermion] (d) -- [fermion] (e),
		};
	\end{feynman}
	\end{tikzpicture}
	+
	\begin{tikzpicture}[baseline=(b)]
	\begin{feynman}[inline = (base.b), horizontal=b to d]
		\vertex [label=180:\({\scriptstyle N,\,p_N}\)](a);
		\vertex [right = of a] (b);
		\vertex [above right = of b, label=360:\({\scriptstyle e^-,\,p_2}\)] (c);
		\vertex [below right = of b] (d);
		\vertex [above right = of d, label=360:\({\scriptstyle \nu_e,\,p_\nu}\)] (e);
		\vertex [below right = of d, label=360:\({\scriptstyle e^+,\,p_1}\)] (f);
		\diagram*{
		(a) -- [fermion] (b) -- [fermion] (c),
		(b) -- [photon] (d),
		(f) -- [fermion] (d) -- [fermion] (e),
		};
	\end{feynman}
	\end{tikzpicture}.
\end{align}
We may note in passing the relative minus sign between the two diagrams 
from the exchange of the positions of $\nu_e$ and $e^-$.
The amplitude square after the spin average is computed as
\begin{align}
	\abs{\mathcal{M}(N \to e^+ e^- \nu_e)}^2 = \frac{g^4}{2m_W^4} \abs{U_{e N}}^2
	\left[4 \sin^4\theta_W (p_2 \cdot p_N) (p_1 \cdot p_\nu)
	+ \left(1+2\sin^2\theta_W\right)^2 (p_2 \cdot p_\nu) (p_1 \cdot p_N)
	\right].
\end{align}
The rest of the computation proceeds in the same way as the muon mixing case.
The decay rate is given by
\begin{align}
	\Gamma(N \to e^+ e^- \nu_e)
	&= 
	\frac{1}{(2\pi)^5}\frac{1}{64 m_N^3}\int_0^{m_N^2} dm_{2\nu}^2
	\int_{0}^{m_N^2 - m_{2\nu}^2} dm_{1\nu}^2
	\int_0^{2\pi}d\psi \int_{-1}^{1}d\cos\theta \int_{0}^{2\pi}d\phi \abs{\mathcal{M}}^2,
	\label{eq:decay_distribution_electron}
\end{align}
and after performing all the integrations, it is given by
\begin{align}
	\Gamma(N \to e^+ e^- \nu_e)
	&= \frac{G_F^2 m_N^5}{768\pi^3}\abs{U_{eN}}^2\left(1 + 4\sin^2\theta_W + 8\sin^4\theta_W\right).
\end{align}
In the electron mixing case, if $m_\mu < m_N < m_\pi$,
the HNL produced from the pion decay can also decay as  $N \to \mu e \nu_e$.
However, as we will see, the LSND constraint for this mass range in the electron mixing case
is anyway weaker than the other constraints, 
and thus we may ignore this decay channel in the rest of the analysis.

\section{Neutrino experiments with stopped muon}
\label{sec:constraint}

Armed with the basic formulas, we now derive the current constraint and future sensitivity
of neutrino experiments with stopped muon on the HNL mixing parameter $U_{lN}$. 
Our signal is $e^+e^-$ produced from the HNL decay.
The total event number of such a decay inside a detector is estimated as
\begin{align}
	N^{(i)}_{ee} &= N_i \times \epsilon_\mathrm{det}
	\times \frac{1}{\Gamma_i}\int dE_N
	\frac{d\Gamma(i \to f N )}{dE_N}
	\times
	\frac{L_\mathrm{det}}{\gamma\beta c\tau_{N\to ee\nu}},
	\label{eq:Nee_total}
\end{align}
where $i = \mu$ if the HNL is produced from the stopped muon and $i = \pi$ if produced from the stopped pion,
and $f$ represents SM particles depending on the decay channel of $i$.
In this expression, 
$N_i$ is the total number of $i$ that decays at the HNL production point, $\epsilon_\mathrm{det}$ is the probability
that a single HNL produced from $i$ passes through the detector, 
and $\Gamma_i$ is the total decay width of $i$.
The relativistic factor $\gamma \beta = (E_N^2 - m_N^2)^{1/2}/m_N$ with $E_N$
being the energy of the HNL, $L_\mathrm{det}$ denotes the length of the detector,
and $\tau_{N \to ee\nu}$ is the decay lifetime of the HNL to a pair of $e^+e^-$ at rest. The signal rate has no $N$ total width $\Gamma_N$ dependence so long as the long lifetime limit is valid, as $1/c\tau_N\times{\rm Br}(N\rightarrow ee\nu)=1/c\tau_{N\rightarrow ee\nu}$.
Note that $\gamma \beta < m_\mu/m_N$ for the HNL produced from the stopped muon.
This enhances the sensitivity to the HNL compared to neutrino experiments with pions and muons decay-in-flight,
such as the MiniBooNE experiment~\cite{MiniBooNE:2008hfu,MiniBooNE:2008paa}.

Of course, experiments impose cuts on events to eliminate backgrounds, and hence we
compute the efficiency $\epsilon^{(i)}$ that the $e^+e^-$ event from the HNL decay passes through the cuts.
For this purpose, we perform a Monte Carlo simulation to generate the distribution of the energy $E_{\pm}$
and the angle $\cos\theta_{\pm}$ \footnote{with respect to the direction of the production point to the center of the detector} 
of the $e^\pm$ in the laboratory frame.
More specifically, we generate the HNL energy, $E_N$, from the differential decay rates
of $\mu$ and $\pi$ computed in Sec.~\ref{subsec:production} , 
and the energy of $e^\pm$ in the $N$-rest frame $E_\pm^N$, as well as the Euler angle ($\psi, \theta, \phi$)
from Eqs.~\eqref{eq:decay_distribution_muon} and~\eqref{eq:decay_distribution_electron}.
We then perform the Lorentz transformation to obtain $E_{\pm}$ and $\cos\theta_{\pm}$ in the laboratory frame.
The HNL event number after imposing the cut is given by
\begin{align}
	N_{ee} = N_{ee}^{(\mu)} \times \epsilon^{(\mu)} + N_{ee}^{(\pi)} \times \epsilon^{(\pi)}.
\end{align}
In the following, we study the current constraint on $U_{lN}$ from the LSND experiment~\cite{LSND:1996jxj},
and the future sensitivity of the PIP2-BD experiment~\cite{Pellico:2022dju,Toups:2022yxs}, respectively.

\subsection{New constraint: LSND}
\label{subsec:LSND}
Here we derive the current constraint on the mixing parameter $U_{lN}$ from the LSND experiment.
The Liquid Scintillator Neutrino Detector (LSND) experiment~\cite{LSND:1996jxj} was located 
at the Los Alamos Neutron Science Center (LANSCE).
An 800\,MeV proton beam is delivered onto the water target  (and other materials such as tungsten), 
producing a large flux of pions dumped at the copper beam stop.
Most of $\pi^-$ are absorbed by the materials and hence the abundance of $\mu^-$ is suppressed.
The main source of the neutrino flux is the decay of stopped $\pi^+$ and $\mu^+$.
The detector is located $\sim 30\,\mathrm{m}$ downstream and is well-shielded.
In the case of our interest, the HNL can be produced from the decay of stopped $\mu^+$ and $\pi^+$.
Once produced, a small portion of the HNL travels downstream and decays inside the detector,
providing $e^+e^-$ signatures.

The LSND experiment did not perform any search on the $e^+e^-$ events. 
Therefore, we may use the single $e$ event search in~\cite{LSND:2001akn} to put a constraint on the HNL.
The number of $\nu_e$ that passes through the detector is $\simeq 3\times 10^{19}$ in total~\cite{LSND:2001akn}.
Since $\nu_e$ is dominantly produced from stopped $\mu^+$,
which is dominantly produced from stopped $\pi^+$, we can set
$N_\mu \times \epsilon_\mathrm{det} = N_\pi \times \epsilon_\mathrm{det} = 3\times 10^{19}$.
Therefore we obtain the total number of the HNL decay events inside the detector originating from $\mu^+$ as
\begin{align}
	N_{ee}^{(\mu)} \simeq
	\begin{cases}
	\displaystyle 1.5\times 10^{5}\left(\frac{\abs{U_{\mu N}}^2}{10^{-6}}\right)^2 
	\left(\frac{m_N}{m_\mu}\right)^6 \left(1-\frac{m_N}{m_\mu}\right)^4 
	\left(1 + \frac{4m_N}{m_\mu} + \frac{m_N^2}{m_\mu^2}\right) 
	& :\mu\mathchar`-\mathrm{mixing}, \vspace{1mm}\\
	\displaystyle 8.1\times 10^{5}\left(\frac{\abs{U_{eN}}^2}{10^{-6}}\right)^2
	\left(\frac{m_N}{m_\mu}\right)^6 \left(1-\frac{m_N}{m_\mu}\right)^4 
	\left(1 + \frac{4m_N}{m_\mu} + \frac{m_N^2}{m_\mu^2}\right) & :e\mathchar`-\mathrm{mixing},
	\end{cases}
\end{align}
and that originating from $\pi^+$ as
\begin{align}
	N_{ee}^{(\pi)} \simeq
	\begin{cases}
	\displaystyle 
	8.7\times 10^{4} \left(\frac{\abs{U_{\mu N}}^2}{10^{-6}}\right)^2 \left(\frac{m_N}{m_\mu}\right)^6 
	\frac{m_\pi\left(
	(m_\mu^2 + m_N^2)m_\pi^2 - (m_\mu^2 + m_N^2)^2 + 4m_\mu^2m_N^2
	\right)}{m_\mu(m_\pi^2-m_\mu^2)^2}
	& :\mu\mathchar`-\mathrm{mixing}, \vspace{1mm}\\
	\displaystyle 4.1\times 10^{5}\left(\frac{\abs{U_{eN}}^2}{10^{-6}}\right)^2
	\left(\frac{m_N}{m_\mu}\right)^7 
	\frac{m_N m_\pi(m_\pi^2 - m_N^2)}{(m_\pi^2 - m_\mu^2)^2} & :e\mathchar`-\mathrm{mixing},
	\end{cases}
\end{align}
where we take $L_\mathrm{det} = 7.6\,\mathrm{m}$ following the fiducial cut specified in~\cite{LSND:2001akn}.
The total number of the HNL decay events is thus sizable, and we can indeed put a stronger 
constraint on the mixing parameter than the other experiments for $m_N < m_\mu$, 
especially in the muon mixing case, as we will see.

\begin{figure}[t]
	\centering
 	\includegraphics[width=0.495\linewidth]{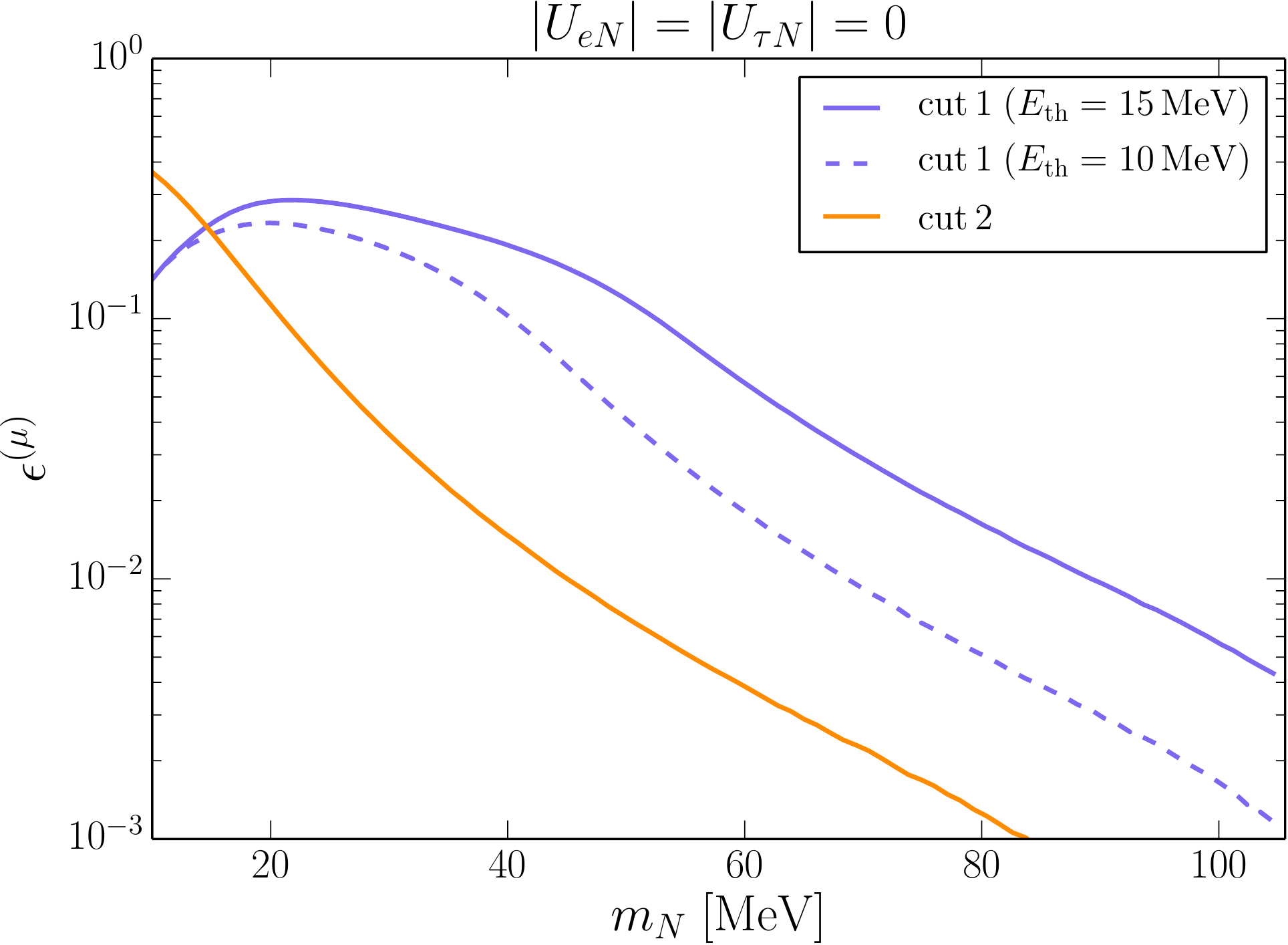}
 	\includegraphics[width=0.495\linewidth]{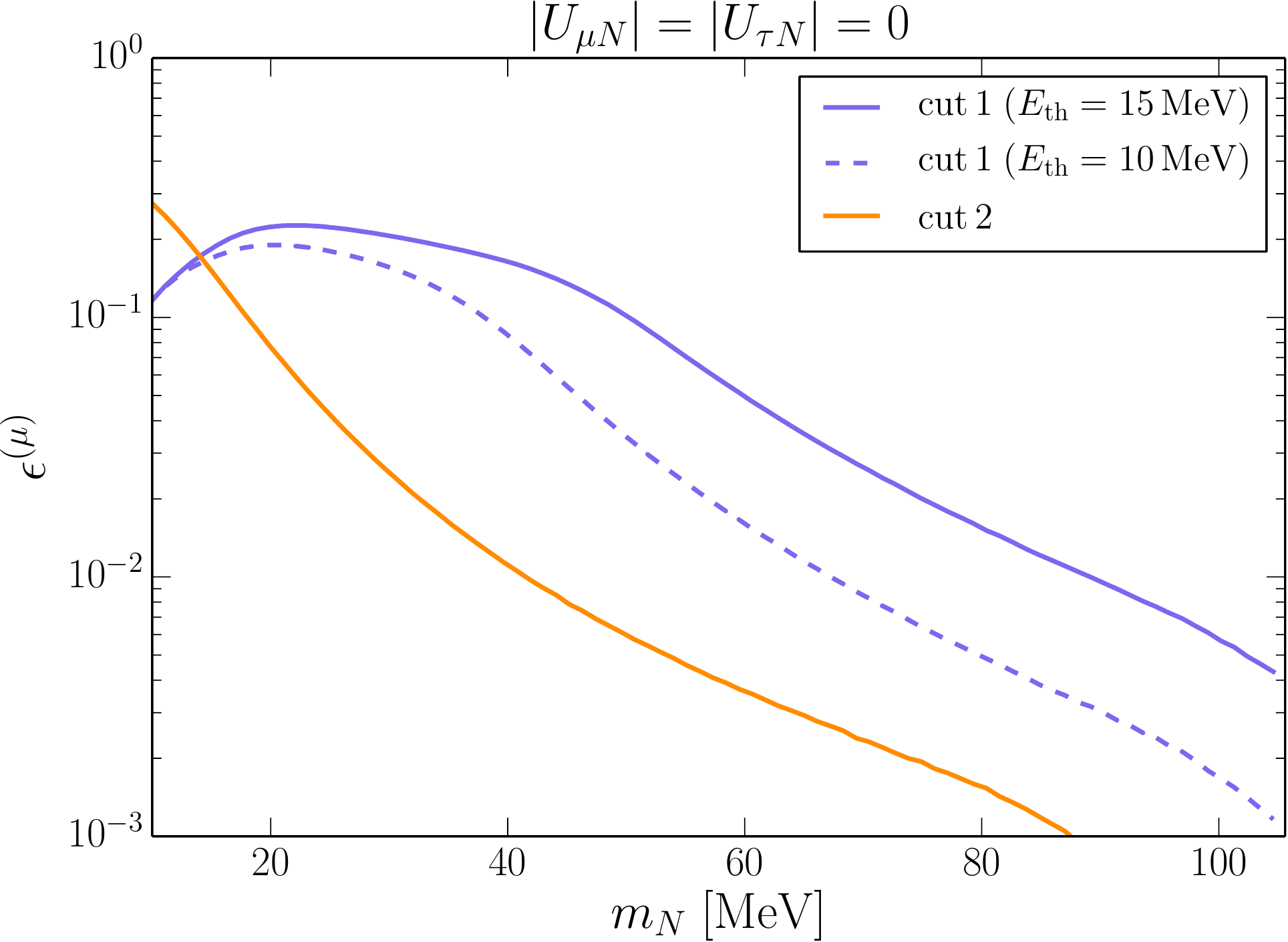}
	\caption{
	The efficiency of the $e^+e^-$ events passing the cuts~1 and~2 with the HNL
	originating from the muon decay. 
	The total efficiency $\epsilon^{(\mu)}$ is the sum of the cuts~1 and~2.
	\emph{Left}: the muon mixing case. 
	\emph{Right}: the electron mixing case.}
	\label{fig:cut_efficiency_LSND}
\end{figure}

Our signal mimics the single $e$ event in~\cite{LSND:2001akn} if either 
(1) one $e^\pm$ has small energy so that it is not detected, while the other $e^\mp$ has enough energy for the detection, 
or (2) $e^+$ and $e^-$ are collinear enough so that they cannot be separated within the detector resolution.
Note that $e^+$ and $e^-$ are produced at the same time in our case
so that it is excluded by neither past nor future activity cuts imposed in~\cite{LSND:2001akn}.
We thus impose two alternative cuts on our events as
\begin{align}
	\begin{cases}
	~\mathrm{cut}~1:~18\,\mathrm{MeV} < E_\pm < 50\,\mathrm{MeV},~E_\mp < E_\mathrm{th},
	~\cos\theta_\pm > 0.9,~\cos\theta_\mp < 0.9, \vspace{2mm}\\
	~\mathrm{cut}~2:~18\,\mathrm{MeV} < E_+ + E_- < 50\,\mathrm{MeV},~\cos\theta_\pm > 0.9.
	\end{cases}
\end{align}
Cut 1 captures the $e^+e^-$ events mimicking single-$e$ events through one of them being soft, and cut 2 captures when they are being colinear.
The events passing either cut 1 or cut 2 will show up 
in the $\nu-e$ scattering data sample of the LSND. 
Here the cut on the angle originates from the requirement that the observed event
has the direction consistent with the neutrino from the production point,
and $E_\mathrm{th}$ is the threshold energy of the $e^\pm$ detection.
We take $E_\mathrm{th} = 15\,\mathrm{MeV}$ as there are huge backgrounds of gamma rays 
in this energy range from, \emph{e.g.},  Carbon excited states and thus 
the analysis is expected to be insensitive to this region. 
We also show the result with $E_\mathrm{th} = 10\,\mathrm{MeV}$ as an even more conservative choice
of the threshold energy. 
In Fig.~\ref{fig:cut_efficiency_LSND}, we show the efficiency of the muon decay channel
$\epsilon^{(\mu)}$ computed by our MC simulation. 
As one can see, the cut~1 dominates for $m_N \gtrsim 20\,\mathrm{MeV}$, while the cut~2 is more important at
the lower mass range as the lighter HNL is more boosted in the forward direction.

\begin{figure}[t]
	\centering
 	\includegraphics[width=0.495\linewidth]{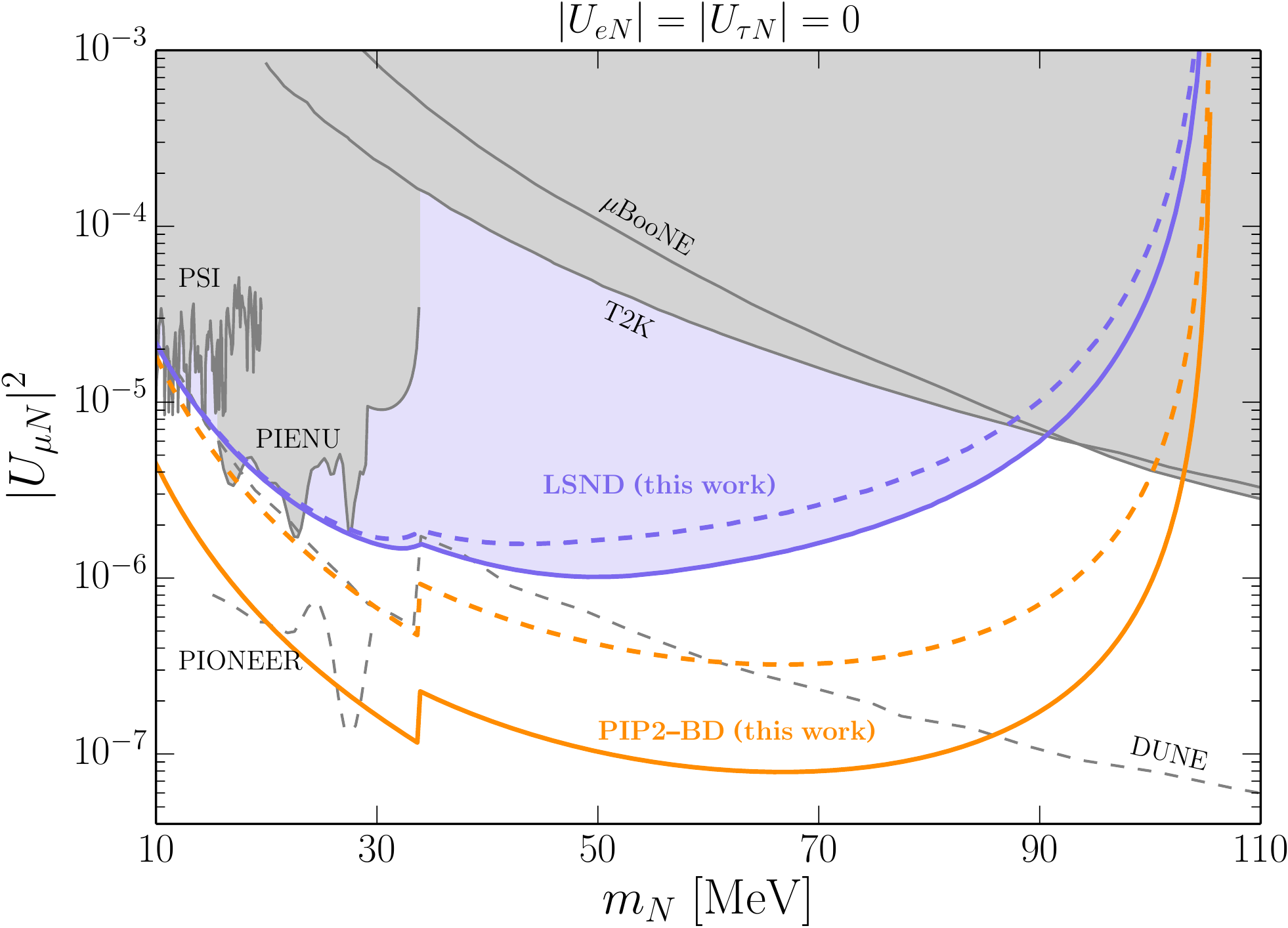}
	\includegraphics[width=0.495\linewidth]{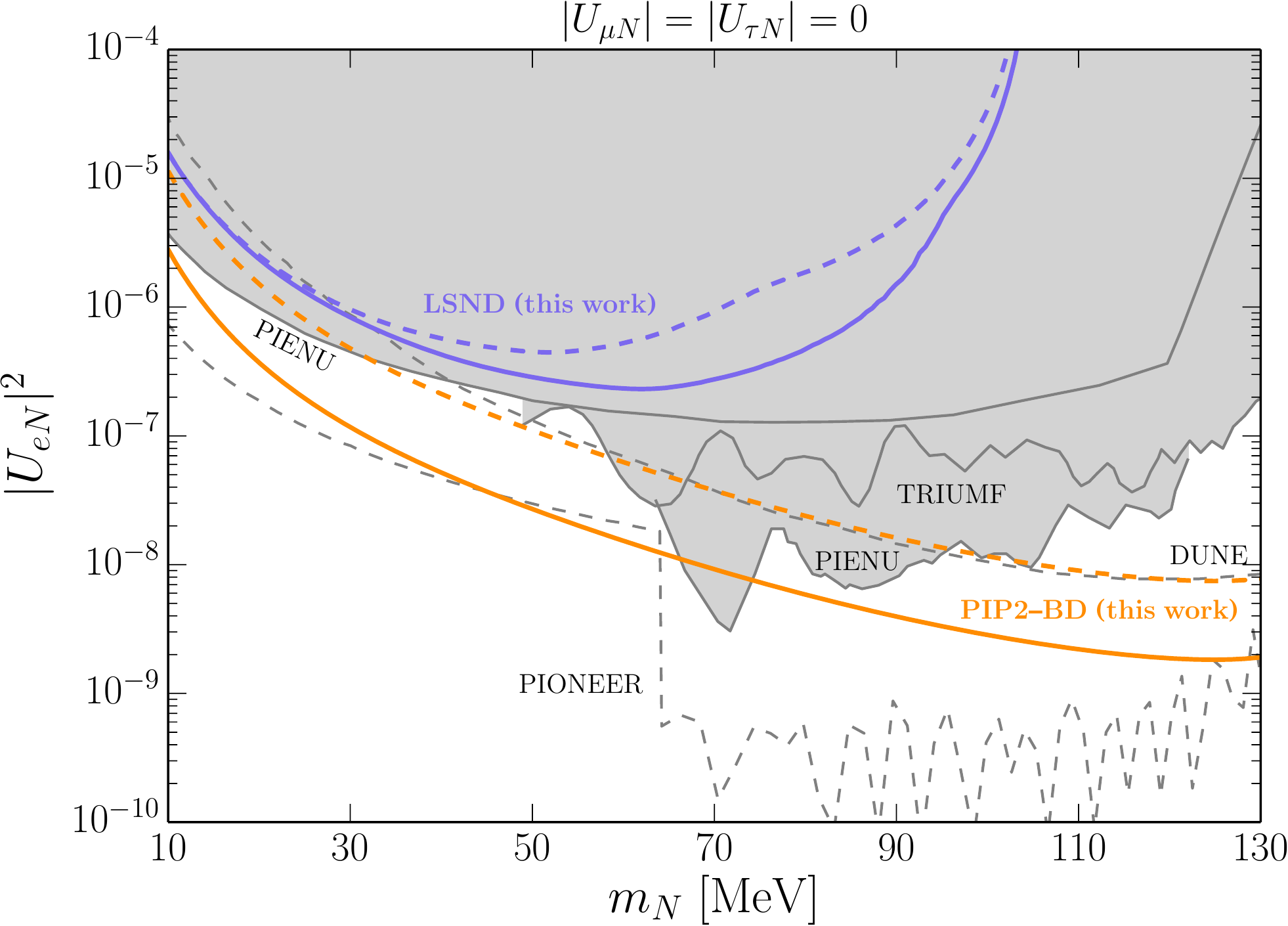}
	\caption{
	The constraints on $\vert U_{l N}\vert^2$ from the LSND experiment
	together with the previous constraints (all at 90\% C.L.) from 
	PSI~\cite{Daum:1987bg}, TRIUMF~\cite{Britton:1992xv},
	PIENU~\cite{PIENU:2017wbj,PIENU:2019usb,Bryman:2019bjg},
	T2K~\cite{T2K:2019jwa,Arguelles:2021dqn}, 
	and $\mu$BooNE~\cite{MicroBooNE:2021usw,Kelly:2021xbv},
	with the data adapted from \texttt{Heavy-Neutrino-Limits}~\cite{Fernandez-Martinez:2023phj}.
	The solid blue line corresponds to the summation of cut~1 with $E_\mathrm{th} = 15\,\mathrm{MeV}$
	and cut~2,
	while the dashed blue line corresponds to that with $E_\mathrm{th} = 10\,\mathrm{MeV}$.
	The future sensitivity of PIP2-BD is also shown,
	where the solid (dashed) orange line corresponds to 3 (50) events of the HNL decay inside the detector,
	together with expected sensitivities of DUNE~\cite{Berryman:2019dme} 
	and PIONEER~\cite{PIONEER:2022yag,PIONEER:2022alm}.
	\emph{Left}: the muon mixing case. \emph{Right}: the electron mixing case.
	The kink around $m_N \simeq 34\,\mathrm{MeV}$ in the muon mixing case
	comes from the pion contribution. 
    In the electron mixing case, the pion contribution dominates over the muon contribution for 
	$m_N \gtrsim 40\,\mathrm{MeV}$.
 }
	\label{fig:mixing_parameter}
\end{figure}

To derive the constraint, we may follow the procedure used in~\cite{LSND:2001akn} 
to put an upper bound on the neutrino magnetic dipole moments. 
The LSND experiment observed in total 242 single $e$ events 
(including all the $\nu_e$, $\nu_\mu$, and $\bar{\nu}_\mu$ initiated signals) while 229 events are expected within the SM.
With the systematic error included, we impose at 90\% C.L.
\begin{align}
	N_{ee} = N_{ee}^{(\mu)} \times \epsilon^{(\mu)} + N_{ee}^{(\pi)} \times \epsilon^{(\pi)} < 55.
\end{align}
In Fig.~\ref{fig:mixing_parameter}, we show the new constraint derived in this study from the LSND (with the future sensitivity
of PIP2-BD, which we describe in the next section) and the previous constraints from PSI~\cite{Daum:1987bg}, 
TRIUMF~\cite{Britton:1992xv}, PIENU~\cite{PIENU:2017wbj,PIENU:2019usb,Bryman:2019bjg},
T2K~\cite{T2K:2019jwa,Arguelles:2021dqn}, and $\mu$BooNE~\cite{MicroBooNE:2021usw,Kelly:2021xbv}.
The solid blue line corresponds to $E_\mathrm{th} = 15\,\mathrm{MeV}$ 
while the dashed blue line does to $10\,\mathrm{MeV}$.
As one can see, in the muon mixing case, our constraint is better than the previous ones by
more than one order of magnitude in terms of $\vert U_{\mu N}\vert^2$ 
for $35\,\mathrm{MeV} \lesssim m_N \lesssim 70\,\mathrm{MeV}$.
In the electron mixing case, the current constraint from LSND is weaker than but close to the constraint from PIENU.

Here we comment on the JSNS$^{2}$ experiment 
at J-PARC~\cite{Ajimura:2017fld,JSNS2:2021hyk}.
JSNS$^{2}$ aims at measuring the neutrino oscillation $\bar{\nu}_\mu \to \bar{\nu}_e$ with the neutrinos
produced from stopped pions and muons created by a 3\,GeV proton beam, 
with the lower duty factor than LSND and the Gd-loaded 
liquid scintillator detector. Thus, JSNS$^{2}$ provides a direct test of 
the LSND neutrino oscillation results~\cite{LSND:2001aii}.
In our case, the HNL decay would be indistinguishable from the single $e$ events (without the final state neutron)
as the JSNS$^{2}$ detector is not sensitive to the angular separation of $e^+$ and $e^-$~\cite{Jordan:2018gcd}.
The current data is $1.45\times 10^{22}$ POT, 13\,\% 
of the designed total POT~\cite{Maruyama:2022juu}, compared to $1.8\times 10^{23}$ POT at LSND
with a larger volume of the detector.
Still, once enough data is accumulated, it would be interesting to study the sensitivity of JSNS${}^{2}$
on the HNL. In particular, JSNS${}^{2}$ can be sensitive to neutrinos from Kaon decay-at-rest, due to higher proton
beam energy, which may allow us to go beyond the muon and pion mass thresholds.

Before closing this subsection, we briefly comment on cosmological constraints on the mixing parameter.
Indeed, in the parameter region of our interest, the HNL decay rate into the SM particles is small so that
it may decay after the BBN, disturbing the consistency 
between the theory and observation~\cite{Sarkar:1995dd,Dolgov:2000jw,Ruchayskiy:2012si,Hufnagel:2017dgo,Boyarsky:2020dzc}.
However, this can be avoided if, \emph{e.g.}, the HNL decays into invisible particles before the BBN.
As long as the HNL decay length is longer than $\sim \mathcal{O}(10\,\mathrm{m})$, this additional decay channel
does not affect our analysis.
In this sense, our constraint is independent of the one from BBN.

\subsection{Future sensitivitity: PIP2-BD}
\label{subsec:future_sensitivity}

In this subsection, we derive the future sensitivity of 
the PIP2-BD experiment~\cite{Pellico:2022dju,Toups:2022yxs}.
The Proton Improvement Project {\romantwo} (PIP-{\romantwo}) is 
a major upgrade of the accelerator complex at Fermilab
to meet the requirement of hosting the Deep Underground Neutrino Experiment (DUNE)~\cite{DUNE:2020lwj,DUNE:2020ypp}.
On top of its main purpose, PIP-{\romantwo} has the flexibility of supporting multiple experiments, 
and a beam dump facility, PIP2-BD, was proposed 
as a candidate experiment to study sub-GeV dark sectors.

In the case of our interest, PIP2-BD can probe the HNL in the same way as LSND.
The HNL is produced from stopped pions and muons at the target.
We may assume five years of physics run with the baseline PAR option~\cite{Pellico:2022dju,Toups:2022yxs},
which results in an 800\,MeV proton beam with $1.2\times 10^{23}$ POT.
We take the formation rate of stopped $\pi^+$ (and hence $\mu^+$) per proton to be 0.1, 
as PIP2-BD is designed to use a lighter target that has a larger formation rate than Hg.
For the mercury target, for instance, the COHERENT experiment reported a formation rate of $(9.0\pm 0.9)\times 10^{-2}$
with slightly higher proton beam energy~\cite{COHERENT:2020iec}.
This fixes the parameters in Eq.~\eqref{eq:Nee_total} as $N_\pi = N_\mu = 1.2\times 10^{22}$.
We assume that the active volume of the detector is cylindrical in shape with 4.5\,m in height and 4.5\,m in diameter, 
located 18\,m away from the HNL production point.
We then estimate from the solid angle coverage that $\epsilon_\mathrm{det} \simeq 3.9\times 10^{-3}$.
The total event number of the HNL decay originating from $\mu^+$ is thus given as
\begin{align}
	N_{ee}^{(\mu)} \simeq
	\begin{cases}
	\displaystyle 1.4\times 10^{5}\left(\frac{\abs{U_{\mu N}}^2}{10^{-6}}\right)^2 
	\left(\frac{m_N}{m_\mu}\right)^6 \left(1-\frac{m_N}{m_\mu}\right)^4 
	\left(1 + \frac{4m_N}{m_\mu} + \frac{m_N^2}{m_\mu^2}\right) 
	& :\mu\mathchar`-\mathrm{mixing}, \vspace{1mm}\\
	\displaystyle 7.5\times 10^{5}\left(\frac{\abs{U_{eN}}^2}{10^{-6}}\right)^2
	\left(\frac{m_N}{m_\mu}\right)^6 \left(1-\frac{m_N}{m_\mu}\right)^4 
	\left(1 + \frac{4m_N}{m_\mu} + \frac{m_N^2}{m_\mu^2}\right) & :e\mathchar`-\mathrm{mixing},
	\end{cases}
\end{align}
and that originating from $\pi^+$ as
\begin{align}
	N_{ee}^{(\pi)} \simeq
	\begin{cases}
	\displaystyle 
	8.0\times 10^{4} \left(\frac{\abs{U_{\mu N}}^2}{10^{-6}}\right)^2 \left(\frac{m_N}{m_\mu}\right)^6 
	\frac{m_\pi\left(
	(m_\mu^2 + m_N^2)m_\pi^2 - (m_\mu^2 + m_N^2)^2 + 4m_\mu^2m_N^2
	\right)}{m_\mu(m_\pi^2-m_\mu^2)^2}
	& :\mu\mathchar`-\mathrm{mixing}, \vspace{1mm}\\
	\displaystyle 3.8\times 10^{5}\left(\frac{\abs{U_{eN}}^2}{10^{-6}}\right)^2
	\left(\frac{m_N}{m_\mu}\right)^7 
	\frac{m_N m_\pi(m_\pi^2 - m_N^2)}{(m_\pi^2 - m_\mu^2)^2} & :e\mathchar`-\mathrm{mixing},
	\end{cases}
\end{align}
where we take $L_\mathrm{det} = 4.5\,\mathrm{m}$.
Without any dedicated study on possible backgrounds at this moment, 
we may simply draw the lines that correspond to 3 and 50 events of the HNL decay inside the detector, 
assuming 75\,\% of event acceptance independent of $m_N$ following~\cite{Toups:2022yxs}.
In Fig.~\ref{fig:mixing_parameter}, we show the future sensitivity of PIP2-BD, together with the expected sensitivities of
DUNE~\cite{Berryman:2019dme} and PIONEER~\cite{PIONEER:2022yag,PIONEER:2022alm}.
The solid orange line corresponds to 3~events while the dashed orange line corresponds to 50~events.
It is clear from the figure that PIP2-BD has the potential of exploring the parameter region 
that is not yet and will not be covered by the other experiments, especially in the muon mixing case.
Here the key difference from LSND is the event acceptance, thus it would be essential to distinguish the $e^+ e^-$ events
from backgrounds, in particular, the single $e$ events from the ordinary neutrinos, to realize this level of sensitivity.

\section{Summary}
\label{sec:summary}

In this paper, we studied the sensitivity of the neutrino experiments with stopped muons and pions
as the source, such as LSND and PIP2-BD, on the HNL.
With the help of the small velocity of the produced HNL as well as a large number of POT,
we find that the LSND constraint on the mixing angle squared is stronger than the previous constraints by 
more than one order of magnitude below the muon mass, if the HNL dominantly mixes with muon neutrinos.
Since LSND does not perform a search for $e^+ e^-$ events, the HNL decay events are compared
with the single $e$ events induced by the SM neutrinos in our analysis.
Therefore, if future experiments such as PIP2-BD can distinguish the $e^+ e^-$ events from backgrounds
induced by the SM neutrinos, the sensitivity will be significantly improved.

Although we have focused on PIP2-BD as a future experiment, 
there are other near-future experiments that are expected to be sensitive to the HNL.
For instance, the Coherent CAPTAIN-Mills (CCM) experiment aims at probing sterile neutrinos 
and sub-GeV dark matters produced at the proton beam dump at LANSCE~\cite{CCM:2021leg}.
Even though the planned number of the total POT as well as the detector volume is smaller than LSND,
CCM may potentially explore a new parameter region of the HNL, 
given that the $e^\pm$ are efficiently resolved by, \emph{e.g.}, the Cherenkov light with their liquid Ar detector.
The COHERENT experiment~\cite{COHERENT:2015mry,COHERENT:2017ipa,COHERENT:2020iec} is another possibility 
once the detector volume is upgraded to be large enough.

The stopped muon and pion sources have been shown to be sensitive probes of sub-100-MeV physics for dark photons, dark scalars, light dark matter, and now for the HNLs. Their sensitivity to axion-like particles (ALPs) could also be interesting. Our initial estimations show that the LSND detector is capable of setting competitive bounds on ALP-photon couplings, which also means that the sensitivity of PIP2-BD to ALPs must also be investigated.

\paragraph{Acknowledgements}

We would like to thank Drs. M. Hostert, P. Huber, W. Louis, T. Maruyama, and R. Tayloe for helpful discussions. 
This work is supported in part by the DOE grant DE-SC0011842. Z.L. and K.L. were supported in part by DOE grant DE-SC0022345. 
The Feynman diagrams in this paper are generated by \texttt{TikZ-Feynman}~\cite{Ellis:2016jkw}. 
The data associated with the figures in this paper can be accessed via \href{https://github.com/ZhenLiuPhys/HNLwStoppedM}{GitHub}.

\appendix

\small
\bibliographystyle{utphys}
\bibliography{ref}
  
\end{document}